\documentclass[a4paper]{article}
\usepackage{amsmath}%
\usepackage{amsfonts,bm}
\usepackage{ifpdf}%
\ifpdf
\usepackage[pdftex]{graphicx}
\usepackage[pdftex, pdfusetitle,colorlinks=false, pdfborder={0 0 0}]{hyperref}%
\DeclareGraphicsExtensions{.png,.pdf}%
\pdfoutput=1 
\pdftrue
\graphicspath{{./figures_pdf/}}%
\usepackage{microtype}%
\else
\usepackage{graphicx}%
\usepackage[colorlinks=false]{hyperref}%
\DeclareGraphicsExtensions{.eps}%
\graphicspath{{./figures/}}%
\fi%
\hypersetup{%
  pdftitle={Sparse Spike Coding : applications of Neuroscience to the processing of natural images.},%
  pdfsubject={version 2},%
  pdfauthor={Laurent Perrinet <Laurent.Perrinet@incm.cnrs-mrs.fr>, INCM/CNRS, 31, ch. Joseph Aiguier, 13402 Marseille Cedex 20, France - http://incm.cnrs-mrs.fr/LaurentPerrinet},%
  pdfkeywords={Neural code, spike-event computation, correlation-based inhibition, Adaptive Matching Pursuit,  Sparse Spike Coding, Competition Optimized Matching Pursuit (COMP).},%
}%
\usepackage[sort&compress]{natbib}%

\newcommand{\sv}{\mathbf{s}} 

\newcommand\la{\leftarrow} %
\newcommand{\eqdef}{\stackrel{\rm def}{=}}%
\newcommand{\seeEq}[1]{Eq.~\ref{eq:#1}}%
\usepackage{units}%
\title{Sparse Spike Coding : applications of Neuroscience to the processing of natural images}%
\author{Laurent U.~Perrinet\thanks{E-mail: \texttt{Laurent.Perrinet@incm.cnrs-mrs.fr}. Further information may be found at \url{http://incm.cnrs-mrs.fr/LaurentPerrinet}, especially \href{http://incm.cnrs-mrs.fr/LaurentPerrinet/SparseSpikeCoding}{supplementary data}  and  \href{http://incm.cnrs-mrs.fr/LaurentPerrinet/Publications/Perrinet08spie}{metadata} about this article, as well as the scripts to reproduce the figures.}\\
Institut de Neurosciences Cognitives de la M\'editerran\'ee (INCM) \\ CNRS / University of Provence\\
31, ch. Joseph Aiguier, 13402 Marseille Cedex 20, France }%
\date{}%
\begin{document}%
\maketitle %
\begin{abstract}%
If modern computers are sometimes superior to humans in some specialized tasks such as playing chess or browsing a large database, they can't beat the efficiency of biological vision for such simple tasks as recognizing and following an object in a complex cluttered background. We present in this paper our attempt at outlining the dynamical, parallel and event-based representation for vision in the architecture of the central nervous system. We will illustrate this on static natural images by showing that in a signal matching framework, a L/LN (linear/non-linear) cascade may efficiently transform a sensory signal into a neural spiking signal  and we will apply this framework to a model retina.  However, this code gets redundant when using an over-complete basis as is necessary for modeling the primary visual cortex: we therefore optimize the efficiency cost by increasing the sparseness of the code. This is implemented by propagating and canceling redundant information using lateral interactions. We compare the efficiency of this representation in terms of compression as the reconstruction quality as a function of the coding length. This will correspond to a modification of the Matching Pursuit algorithm where the ArgMax function is optimized for competition, or Competition Optimized Matching Pursuit (COMP). We will in particular focus on bridging neuroscience and image processing and on the advantages of such an interdisciplinary approach.%
\end{abstract}%
{\bf Keywords}: Neural population coding, decorrelation, spike-event computation, correlation-based inhibition, Sparse Spike Coding, Competition Optimized Matching Pursuit (COMP)%
\section{Introduction: efficient neural representations}%
\label{sec:intro}%
The architecture of modern day computers illustrate how we understand intelligence. But, if they are good at playing chess or at browsing databases, it is clear that computers are far from rivaling with what appears to be more simple aspects of intelligence such as the ones demonstrated in vision. Think for instance as something as simple as recognizing an object in natural conditions, such as while walking in the street. This necessarily involves a network of processes from segmenting its outline, perceiving its global motion, matching its different patterns invariantly to the shading, contrast, angle of view or to occlusions. Actually, while this seems obvious to us, computers cannot perform this task and it is a common practical ``Turing Test" to authenticate humans versus spamming robots by challenging the login upon recognizing for instance warped letters on a noisy background (the so-called \href{http://en.wikipedia.org/wiki/CAPTCHA}{CapTchas}).\\%
As the seat of this processing, the Central Nervous System (CNS) is therefore by its efficiency clearly different from a classical~\citet{Neumann66} computer defined as a sequential Turing-like machine with a few, very rapid Central Processing Units and a finite, adressable memory. Computational Neuroscience is a branch of neuroscience studying specifically the structure and function of computations in the CNS such as the more complex architectures imagined by~\citet{Neumann00}. Numerous successful theories exist to explain the complex dynamics of modern Artificial Neural Networks and how we may use neuro-physiological constraints to build up efficient systems~\citep{Grossberg03} that are ecologically adapted to the statistics of the input~\citep{Atick92}. However, a main challenge involving both neuroscience and computer science is to understand how and for what class of problems the CNS outperforms traditional computers. I am interested in this paper in extracting general principles from the structure of the CNS to derive a better understanding of the neural functions but also to apply these algorithms to signal processing applications.\\%
A fundamental difference of the CNS is the fact that 1) information is distributed in parallel on the different neurons, 2) processes are dynamical and interruptible, 3) information is carried by elementary events, called \emph{spikes} which may be transmitted over long distances. This is well illustrated for the large class of pyramidal neurons of the neocortex. In a simplistic way the more a neuron is excited, the quicker and the more often it will emit spikes, with a typical latency of some milliseconds and a maximum firing frequency of the order of \unit[200]{ms}. Concentrating on local cortical areas (that is in human to the order of some squared centimeters and to a billion neurons), it means that the complexity of some operation will be different on a computer (a few but very rapid CPUs) and a population of neurons (a huge number of slow dynamical event generators). For instance, the complexity of the ArgMax operator (finding the sorted indices from a vector) will increase as $O(Nlog(N))$ with the dimension $N$ of the vector, while if we apply the vector as the activation of a neuronal population, the complexity will not increase with the number $N$ of neurons\footnote{Note that in a noisy environment, the output will be given with a certain temporal precision and that this precision may decrease with $N$.}. In addition, the result is given by the generated  spike list and is interruptible. \\%
In this paper, we will explore how we may apply this class of operators to the processing of natural images by presenting an adaptive Linear/Non-Linear framework and then optimize its efficiency. We will in a first step draw a rationale for using a linear representation by linking it to a probabilistic representaiton under the condition of decorrelation. Then we will derive a linear transform adapted to natural images by constructing a simple pyramidal architecture similar to~\citep{Burt83} and extend it to a Laplacian and Log-Gabor pyramids~\citep{Fischer05a}. We will then in a third section propose that this linear information may be optimally coded by a spike list if we apply a point non-linear operation. At least, we will define an improvment over Matching Pursuit~\citep{Mallat93} by optimizing the efficiency of the ArgMax operator and which finally defines Sparse Spike Coding~\citep{Perrinet02sparse,Perrinet04tauc,Perrinet06}.%
\section{Linear filtering and whitening}%
\label{sec:white}%
A first step in the definition of this algorithm is to explicit the linear operations which are used to transform the input vector into a value representative of the quality of a match. %
Let's define an image as a set of scalar values $\tilde{x}_i$ on a set of positions ${\cal P}$, $i$ being the index of the positions, so that it defines a vector $\tilde{x} \in \mathbb{R}^M$, with $M=card({\cal P})$. As we saw in previous works~\citep{Perrinet04tauc}, the quality of a match between the raw data $\tilde{x}$ with a known image may be linked in a probabilistic framework to the correlation coefficient. In fact, the probability of the signal $\tilde{x}$ knowing the ``shape'' $\tilde{h}$ of the signal to find (see the table Tab.~\ref{tab:linear} for the chosen notation) is:
\begin{eqnarray}
P( \tilde{h} | \tilde{x} ) &=& \frac{1}{P(\tilde{x})} P( \tilde{x} | \tilde{h} ) P(\tilde{h} ) \nonumber\\
&=& \frac{1}{P(\tilde{x})} \frac{1}{(2\pi)^{M/2}} \exp(-\frac{ (\tilde{x} - \tilde{h}){\bm \Sigma}^{-T} (\tilde{x} - \tilde{h})^T }{2}).P( \tilde{h} ) \nonumber\\
\label{eq:log_proba}%
\end{eqnarray}
This is based on the assumption of centered data (that is $E(x)=0$), a Linear Generative Model and a gaussian noise of covariance matrix ${\bm \Sigma} = E(\tilde{x}\tilde{x}^T)$ (See Chapter~2.1.4 of ~\citep{Perrinet06}). In the case where the noise is white (that is that the covariance matrix is a diagonal matrix) and assuming an uniform prior for the scalar value of $h$, this may be simply computed with the correlation coefficient defined by:
\begin{eqnarray}%
\rho = <\frac{h}{\|h\|}, \frac{x}{\|x\|}>   \eqdef \frac{ \sum_{1\leq i \leq M} x_i h_i }{\sqrt{\sum_{1\leq i \leq M} h_i^{2} } \sqrt{\sum_{1\leq i \leq M} x_i^{2} } }%
\label{eq:coco}%
\end{eqnarray}%
\begin{figure}
   \begin{center}
   \begin{tabular}{c}
 \includegraphics[width=.9\linewidth]{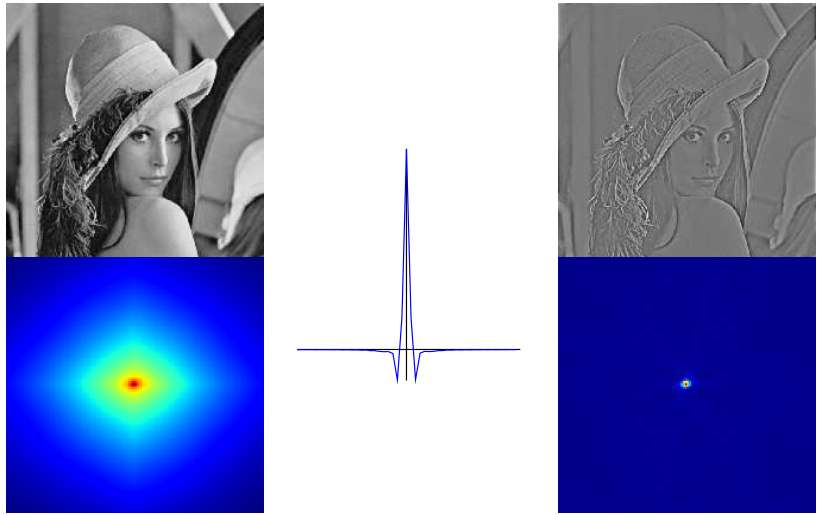}
   \end{tabular}
   \end{center}
   \caption[whitening] 
{ \label{fig:white} \rm{Spatial decorrelation. } \emph{(Top-Left)} Sample raw natural image ($M=512^2$). \emph{(Bottom-Left)} Mean pairwise spatial correlation in a set of $1000$  natural images (Red is 1, blue is zero). It shows the typical decrease in $ \frac{1}{f^2}$ of the power spectrum but also an anisotropy along the vertical and horizontal axis.  \emph{(Middle)} decorrelation filter computed from the methods of~\citep{Atick92} (see text). This profile is similar to the interaction profile of bipolar and horizontal cells in the retina. \emph{(Top-Right)} Whitening of the sample image. \emph{(Bottom-Right)} The mean pairwise spatial correlation of $1000$ whitened natural images is highly peaked at the origin and inferior to $0.05$ elsewhere. As is observed in the LGN, the power spectrum is relatively decorrelated by our pre-processing~\citep{Dan96}.  See script \texttt{experiment\_whitening.py} to reproduce the figure.}
\end{figure}%
It should be noted that $\rho_j$ is the $M^{\rm th}$-dimensional cosinus and that its absolute value is therefore bounded by 1. The value of $\mbox{ArcCos}(\rho_j)$ would therefore give the angle of $x$ with the pattern $h$ and in particular, the angle would be equal (modulo $2\pi$) to zero if and only if $\rho_j=1$ (full correlation), $\pi$ if and only if $\rho_j=-1$ (full anti-correlation) and $\pm\pi/2$ if $\rho_j=0$ (both vectors are orthogonal, there is no correlation). Also, it is independent to the norm of the filters and we assume without loss of generality in the rest that these are normalized to unity.  To achieve this condition, the raw data $\tilde{x}$ has to be preprocessed with a decorrelation filter to achieve a signal $x$ with no mean point-wise correlation\footnote{Of course, this does not achieve necessarily independence as is often stated.}. To define this, we may use for instance  the eigenvalue decomposition (EVD) of the covariance matrix:
\begin{eqnarray}
{\bm \Sigma} = \mathbf{V}\mathbf{D}\mathbf{V}^T \nonumber\\
\label{eq:evd}%
\end{eqnarray}
where $\mathbf{V}$ is a rotation (and thus $\mathbf{V}^{-1}=\mathbf{V}^T$) and $\mathbf{D}$ is a diagonal matrix. This decomposition is similar to that achieved by PCA and may be computed for instance by averaging linear correlations such as is done with the linear Hebbian rule~\citep{Oja82}. In  particular, the columns of matrix  $\mathbf{V}$  contain the eigenvectors and $\mathbf{D}$  is a diagonal matrix of the corresponding eigenvalues. If we set $\mathbf{W}= \mathbf{D}^{-\frac{1}{2}}\mathbf{V}^T$ and $x= \mathbf{W}\tilde{x}$, then
\begin{eqnarray}
E(xx^T) &=&  E(\mathbf{W}\tilde{x} (\mathbf{W}\tilde{x})^T ) \nonumber\\
 &=&  \mathbf{D}^{-\frac{1}{2}}\mathbf{V}^T E(\tilde{x} \tilde{x}^T) (\mathbf{D}^{-\frac{1}{2}}\mathbf{V}^T)^T \nonumber\\
 &=&  \mathbf{D}^{-\frac{1}{2}}\mathbf{V}^T {\bm \Sigma} \mathbf{V}\mathbf{D}^{-\frac{1}{2}} \nonumber\\
 &=&  \mathbf{D}^{-\frac{1}{2}}\mathbf{V}^T \mathbf{V}\mathbf{D}\mathbf{V}^T \mathbf{V}\mathbf{D}^{-\frac{1}{2}} \nonumber\\
 &=&  \mathbf{1}^{M\times M} \nonumber
\end{eqnarray}
We therefore proved that this linear transform de-correlates on average the input data. In practice, we used the power spectrum and its relation to the covariance in translation invariant data such as natural images to compute the whitening filter~\citep{Atick92}. This corresponds then to a filter with a gain proportional to the spatial frequency but with an anisotropy  on the vertical and horizontal axis (see Fig.~\ref{fig:white}).\\%
Thanks to this processing, and only when these hypothesis have been fullfilled, we may in general use the correlation coefficient (see Eq.~\ref{eq:coco}) as a measure related to the probability of a match of the image with a given pattern. The next step is now to define the best patterns to represent images.%
\begin{table}[hbpt]%
\label{tab:linear}%
\begin{center} %
\caption{Matrix notation and denoising Variables}
\begin{tabular}{|l|c|c|}  \hline
Name&Symbol&Description\\\hline\hline %
Pixel positions&${\cal P}$&$ \vec{p} \in {\cal P}, card({\cal P}) = M$ \\ 
Raw image&$\tilde{x}$&$\tilde{x} \in \mathbb{R}^M$, $E(\tilde{x}) =0$ \\ 
Covariance matrix&${\bm \Sigma}$&${\bm \Sigma} \in \mathbb{R}^{M\times M}$\\
Whitening matrix&$\mathbf{ W}$&$\mathbf{ W}\in \mathbb{R}^{M\times M}$\\
Decorrelated image&$x$&$x= \mathbf{W}\tilde{x} \in \mathbb{R}^M$\\
Pattern image&$\tilde{h_j}$&$ h_j \in \mathbb{R}^M, j \in {\cal D}$\\
Overcomplete dictionary&${\cal D}$&$card({\cal D})= N \gg M$\\
Decorrelated pattern image&$h_j$&$h_j= \mathbf{W}\tilde{h_j} \in \mathbb{R}^N$\\
Transform matrix&$\mathbf{H}$&$ \mathbf{H} \in  \mathbb{R}^{N\times M}$\\
Correlation coefficient&$\rho_j$&$\rho_j = \frac{<h_j,x>}{\|h_j\|\|x\|} \in  [ -1,1 ] $\\
\hline\hline%
\end{tabular}
\end{center}
\end{table}
\section{Multiscale representations: the (Golden) Laplacian Pyramid}%
\label{sec:pyramid}%
Multi-scale representations are a popular method to allow for a scale invariant representation. This correspond to repeating basic shapes at different scales and it thus allows that one may easily compute the representation of a scaled image by a simple transformation in the representation space instead of recomputing the whole transform. As a consequence, this representation makes it for instance easier to compute the match of a feature at different scales. It is classically implemented in wavelet transforms but we present here a simple implementation using a recurrent scheme, the Laplacian Pyramid~\citep{Burt83}. This transform has indeed the advantage of being computed by simple down-scaling and up-scaling operations and is easily inverted for the reconstruction of the image. It transforms an image in a list of down-scaled images, or \emph{image pyramid}. Let's define the list $\{ M^k \}$ with $0 \leq k \leq s$ of the sizes of the down-scaled images ($k=0$ corresponds to the ``base'' and $M^0=M$ while $s$ is the level of the smallest image, that is the summit of the pyramid). Typically, such as in wavelets, the size decreases geometrically with an exponent $\gamma$. The most used exponent in image processing is $2$, the pyramid is then called \emph{dyadic}. The corresponding down-scale and up-scale transform from level $k$ to $k+1$ may be defined as ${\cal D}_k$ and  ${\cal U}_k$ respectively. We may therefore define the gaussian pyramid as the recursive transform from the ``base'' of the pyramid to the top as the list of transforms:  %
\begin{eqnarray}
{\cal G} = \{ {\cal D}^k \} \mbox{ with } {\cal D}^k = {\cal D}_0 \circ  \cdots \circ {\cal D}_k
\label{eq:gpyr}%
\end{eqnarray}
\begin{figure}
   \begin{center}
   \begin{tabular}{c}
 \includegraphics[width=.9\linewidth]{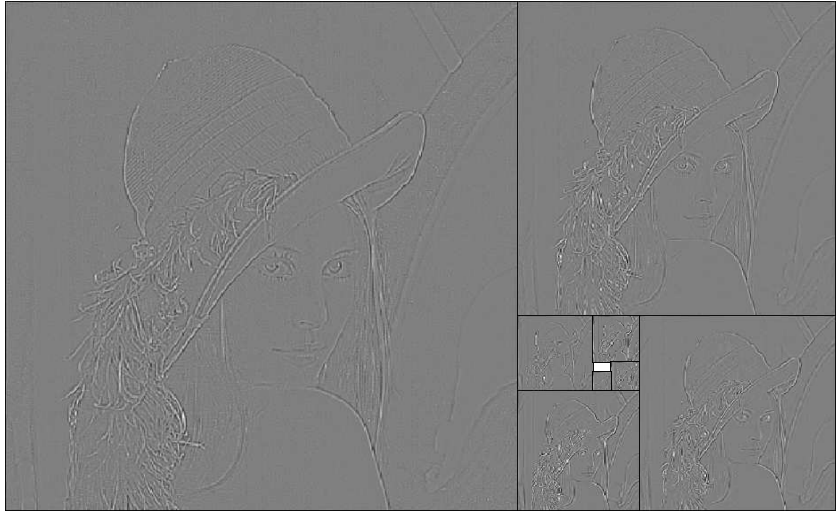}
   \end{tabular}
   \end{center}
   \caption[SC] 
   {\label{fig:golden_pyr} \rm{The Golden Laplacian Pyramid. }%
To represent the edges of the image at different levels, we may use a simple recursive approach constructing progressively a set of images of decreasing sizes, from a base to the summit of a ``pyramid''. Using simple down-scaling and  up-scaling operators we may approximate well a Laplacian operator. This is represented here by stacking images on a ``Golden Rectangle'', that is where the aspect ratio is the golden section $\phi \eqdef \frac{1+\sqrt{5}}{2}$. We present here the base image on the left and the successive levels of the pyramid in a clockwise fashion (for clarity, we stopped at level $8$).  Note that here we also use $\phi^2$ (that is $\phi+1$) as the down-scaling factor so that the resolution of the pyramid images correspond across scales. Note at last that coefficient are very kurtotic: most are near zero, the distribution of coefficients has ``long tails''.    See script \texttt{experiment\_SpikeCoding.py} to reproduce the figure.}
\end{figure} 
This means that a down-scaled version of the image ${\cal D}^k x$ may be obtained by applying all down-scaling transforms sequentially from the base to level $k$. If the elementary operators are linear, the ${\cal G}$ transform is linear. The corresponding filters correspond approximately  to gaussians with increasing radiuses~\citep{Burt83} and the images in the pyramid thus correspond to progressively more blurred versions of the ``base'' image. This transform is usually very fast and is very likely to be implemented by the extended dendritic arbor of neurons\footnote{Note however that in vertebrates, the retinal representation the preferred spatial frequency grows with eccentricity.}.\\%
The Laplacian Pyramid is defined from the Gaussian Pyramid as the pyramid of images constituted by the residual between the image at one scale and the up-scaled image from the upper level. It is therefore mathematically defined as:
\begin{eqnarray}
{\cal L} = \{ {\cal D}^k -({\cal U}_k \circ  {\cal D}^{k+1}) \} \mbox{ with } 0 \leq k  \leq s
\label{eq:lpyr}%
\end{eqnarray}
by defining for clarity that ${\cal D}^{0} = 1$ and  ${\cal D}^{s+1} = 0$. This transform is still linear that is that $\forall x, \forall y, \forall \lambda$,  ${\cal L}(x+y) = {\cal L}x + {\cal L}y$ and ${\cal L}(\lambda x) = \lambda {\cal L}x$. Since every level corresponds to the residual, it is easy to invert. In fact, if we write as ${\cal L}_k x$ the image at level $k$ and ${\cal U}^k = {\cal U}_0 \circ  \cdots \circ {\cal U}_k$, then $\forall x$,
\begin{eqnarray}
\sum_{0 \leq k  \leq s} {\cal U}^k {\cal L}_k x &=& \sum_{0 \leq k  \leq s} {\cal U}^k({\cal D}^k -({\cal U}_k \circ  {\cal D}^{k+1})) x  \nonumber\\
&=& \sum_{0 \leq k  \leq s} {\cal U}^k{\cal D}^k x -  \sum_{0 \leq k  \leq s} {\cal U}^k{\cal U}_k \circ  {\cal D}^{k+1} x \nonumber\\
&=& \sum_{0 \leq k  \leq s} {\cal U}^k{\cal D}^k x -  \sum_{1 \leq k  \leq s+1} {\cal U}^k \circ  {\cal D}^{k} x  = x
\label{eq:lpyr_rec_proof}%
\end{eqnarray}
Therefore the inverse of the Laplacian Pyramid transform is defined as:
\begin{eqnarray}
 {\cal L}^{-1} = \sum_{0 \leq k  \leq s} {\cal U}^k {\cal L}_k
\label{eq:lpyr_rec}%
\end{eqnarray}
The filters corresponding to the different levels of the pyramid (and which are the inverse image of a Dirac pyramid by ${\cal L}^{-1}$) are similar to difference of gaussians (because they are the difference of  two successive  levels of the Gaussian Pyramid). The exponent $\gamma$ will therefore play the important role of the ratio of the the radiuses of the Gaussians. We choose here the exponent to be equal to the golden number $\gamma = \phi \eqdef \frac{1+\sqrt{5}}{2} \approx 1.618033$ for two reasons. First, it corresponds to a value which approximates well a Laplacian-of-Gaussians with a Difference of Gaussians as is implemented here. Second, it allows to construct a natural representation of the whole pyramid in a full Golden Rectangle (see Fig.~\ref{fig:golden_pyr}) where the resolution of each image will be constant.\\%
Note the following properties of the pyramid:
\begin{itemize}
\item the over-completeness is equal to $\sum_{0 \leq k  \leq s} \frac{1}{\gamma^{2k}}  \approx \frac{1}{1-\gamma^{-2}}$ so that it is equal to $\frac{1}{1-\phi^{-2}} = \frac{\phi}{\phi - \phi^{-1}} = \phi$ which is indeed the area of the Golden Rectangle compared to the area of the image. It is slightly higher than for a dyadic pyramid (indeed $\frac{1}{1-2^{-2}}=\frac{4}{3} \approx 1.333 < \phi$). 
\item  since this linear transform is over-complete, there may exist non zero pyramids which inverse image is null (that is $\exists L\neq 0$ such that $ {\cal L}^{-1} L = 0$) but this pyramids are not accessible from any non-null image.
\item one may also implement a simple ``Golden Pyramid'' using the Fourier transform, and one may observe that in both cases, the filters corresponds to localized filters in the frequency space. The whitening (see Sec.~\ref{sec:white}) has an approximately scalar effect that  corresponds to an equalization of the variances of the coefficients to natural images at the different spatial frequencies.
\item Finally, once  the obtained filters are normalized, the coefficients will correspond to the correlation coefficients of the image with edge detectors at different scales as defined in Eq.~\ref{eq:coco}. The coefficients will therefore as in wavelet analysis correspond to the local Lipschitz coefficients of the image~\citep{Perrinet03ieee}. When ordered by decreasing absolute values they will  correspond to features of decreasing singularities, from a pure singularity, to a smooth transition (as a ramp of luminosity).%
\end{itemize}%
\begin{table}[hbpt]%
\label{tab:pyramid}%
\begin{center} %
\caption{Notations used for the Laplacian Pyramid}
\begin{tabular}{|l|c|c|}  \hline
Name&Symbol&Description\\\hline\hline %
sizes of the down-scaled images&$\{ M^k \}$& $0 \leq k \leq s$ \\ 
Down-scale operator&${\cal D}_k$ & from level $k$ to $k+1$ \\ 
Up-scale operator& ${\cal U}_k$& from level $k$ to $k+1$ \\
Full Down-scale operator&${\cal D}^k$ & ${\cal D}^{0} = 1$ and  ${\cal D}^{s+1} = 0$ \\ 
Full Up-scale operator& ${\cal U}^k$& \\
Gaussian Pyramid&${\cal G}$&\\
Laplacian Pyramid&${\cal L}$ & ${\cal L} = \{ {\cal L}_k\}$ with  $0 \leq k \leq s$\\
\hline\hline%
\end{tabular}
\end{center}%
\vspace*{-.4cm}
\end{table}
\section{Spike Coding}%
\label{sec:SC}%
Now that we defined a linear transform which is suitable for natural images by associating the whitening filters and the Laplacian Pyramid, we wish  to transmit this information efficiently using neurons. As we saw in the previous section, the higher coefficients correspond to more singular features and therefore to more informative content. By using Integrate-and-Fire neurons, it is therefore natural that we may associate to every coefficient  of the pyramid applied to the image a single neuron. For the linear Leaky-IF,  if we associate a driving current to each value $\rho_j$ (with $0 \leq j \leq N$, as noted in Tab.~\ref{tab:linear}) it will will elicit spikes with latencies~\citep{Perrinet03ieee}:
 \begin{eqnarray}
\lambda_j = \tau \log\frac{1}{1 - \theta . g_j(\rho_j)}
\label{eq:lif}
\end{eqnarray}%
where $\tau$ is the characteristic time constant, $\theta$ is the neuron's threshold and $g_j$ is a monotonously increasing function of $\rho_j$ corresponding to the transformation of the linear value into the driving current. By this architecture, since the relation in Eq.~\ref{eq:lif} is monotonously increasing, one implements a simple ArgMax operator where the output is the index of the neurons corresponding to the ordered list of output spikes.\\%
However, one may observe that for some linear transforms, the distribution of correlation coefficients may be not similar for all $j$. This is contradictory with the fact that spikes are similar across the CNS since it would mean that the probability of the coefficient underlying the emission of a spike is not uniform. To optimize the efficiency of the ArgMax operator, one has therefore to ensure that  one optimizes the entropy of the index of output spikes and therefore of the driving current. This may be ensured by modifying the functions $g_j$ so that:
\begin{enumerate}
\item for all $j$, the distributions of $g_j(\rho_j)$ are similar,
\item allow that this overall distribution has a shape adapted to the spiking mechanism (for instance by using Eq.~\ref{eq:lif}).
\end{enumerate}
The second point ---finding a global non-linearity $g$--- will be out of scope of this paper, and we will for the sake of generality only ensure that  we find functions $f_j$ (with $g_j =  g \circ f_j$) such that the variables $z_j = f_j(\rho_j)$ are uniformly distributed.\\%
This condition is easily performed by operating a point non-linearity on the different variables $\rho_j$ based on the statistics of natural images~\citep{Atick92}. This method is similar to histogram equalization in image processing and provides an output with maximum entropy for a bounded output: it therefore optimizes the coding efficiency of the representation in terms of compression~\citep{Hateren93} or dually the minimization of intrinsic noise~\citep{Srinivasan82}. It may be easily derived from the probability $P$ of variable $\rho_j$ (bounded in absolute value by $1$) by choosing the non-linearity as the cumulative function
\begin{equation}
f_j(\rho_j)=\int_{-1}^{\rho_j} dP(\rho)%
\label{eq:laughlin}%
\end{equation}
where the symbol $dP(x) = P_X(x) dx$ will here denote in general the probability distribution function (pdf) for the random variable $X$. This process has been observed in a variety of species and is for instance perfectly illustrated in the salamander~\citep{Laughlin81}. It may evolve dynamically to slowly adapt to varying changes in luminances, such as when the light diminishes at dawn but also to some more elaborated schemes within a map~\citep{Hosoya05}. As in ``ideal democracies'' where all neurons are ``equal'', this process has to be dynamically updated over some characteristic period so as to achieve optimum balance. As a consequence, since for all $j$, the pdf of $z_j = f_j(\rho_j)$ is uniform and that sources are independent, it may be considered as a random vector drawn from an uniform distribution in $[0, 1]$. Knowing the different spike generation mechanisms which are similar in that class of neurons, every vector $\{ \rho_j \}$ will thus generate a list of spikes $\{ j(1), j(2), \ldots \}$ (with corresponding latencies) where no information is carried \emph{a priori} in the latency pattern but all is in the relative timing across neurons.\\%
We coded the signal in a spike volley, but how can this spike list be ``decoded'', especially if it is conducted over some distance and therefore with an additional latency? In the case of transient signals, since we coded the vector $\{ \rho_j \}$ using the homeostatic constraint from \seeEq{laughlin}, we may retrieve the analog values from the order of firing neurons in the spike list. In fact, knowing the ``address'' of the fiber $j(1)$ corresponding to the first spike to arrive at the receiver end, we may infer that it has been produced by a value in the highest quantile of $P(\rho_{j(1)})$ on the emitting side. We may therefore decode the corresponding value with the best estimate $\hat{\rho}_{j(1)} = f_{j(1)}^{-1}(\frac{1}{N})$ where $N$ is the total number of neurons. This is also true for the following spikes and if we write as $z_{j(k)}=\frac{k}{N}$ the relative rank of the spike (that is neuron $j(k)$ fired at rank $k$), we can reconstruct the corresponding value as%
\begin{equation}%
\hat{\rho}_{j(k)}=f^{-1}_{j(k)}(1- z_{j(k)} )%
\label{eq:mod}%
\end{equation}%
\begin{figure}
   \begin{center}
   \begin{tabular}{c}
 \includegraphics[width=\linewidth]{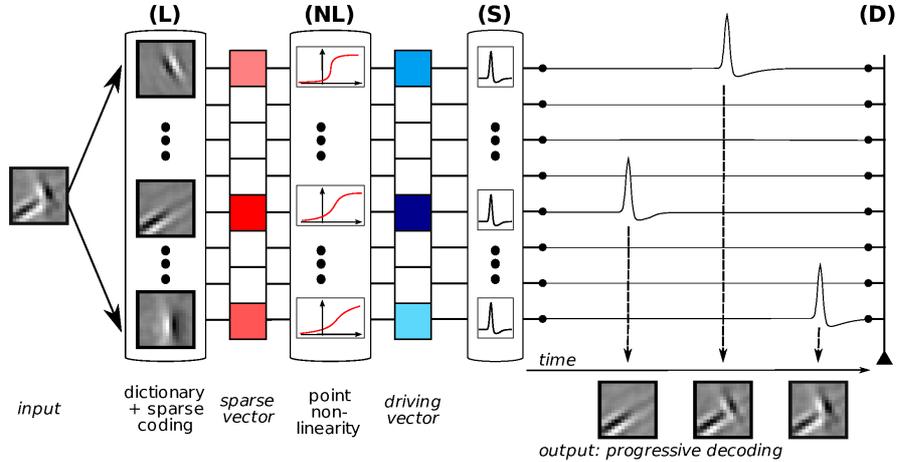}
   \end{tabular}
   \end{center}
   \caption[example] 
   { \label{fig:SSC} \rm{Spike Coding of natural images. }%
We did build here a simple framework of pyramidal neurons illustrating the efficiency of neural architectures compared to classical computer architectures. We show here how a bundle of L-NL neurons~\citep{Carandini97,Carandini05} tuned by a simple homeostatic mechanism allow to transfer a transient information, such as an image, using spikes. (L) The signal to be coded, for instance the match $\rho_j$ of  an image patch (the tiger on the left bottom) with a set of filters (edge-like images), may be considered as a stochastic vector defined by the probability distribution function (pdf) of the values $\rho_j$ to be represented. (NL) By using the cumulative function as a point non-linearity $f_j$, one ensures that the probability of $z_j = f_j(\rho_j)$  is uniform, that is that the entropy is maximal. This non-linearity in the L-NL neuron implements a homeostasis that is controlled only by the time constant with which the cumulative probability function $f_j$ is computed (typically $10^4$ image patches in our case). (S) Any instance of the signal may then be coded by a volley of spikes: a higher value corresponds to a shorter latency and a higher frequency. (D) Inversely, for any spike events vector, one may estimate the value from the firing frequency, the latency. We may simply use the ordering of the spikes since the rank provides an estimate of the quantile in the probability distribution function thanks to the equalization. Using the inverse of $f_j$ one retrieves the value in feature space so that this volley of spikes is decoded (or directly transformed) thanks to the relative timing of the spikes using the modulation (see \seeEq{mod}). This builds a robust information channel where information is solely carried by spikes as binary events. Given this model, the goal of this work is to find the most efficient architecture to code natural images and in particular to define a coding cost and to derive efficient compression algorithms. Note that this scheme is similar to the N-NL scheme but that instead of generating a Poisson point process, we use the the exact timing. This is allowed by the point non-linearity which permits to code the value by the timing and not the firing frequency.}
\end{figure} 
This corresponds to a generalized rank coding scheme~\citep{Perrinet99,Perrinet01}. First, it loses the information on the absolute latency of the spike train which is giving the maximal value of the input vector. This has the particular advantage of making this code invariant to contrast (up to a fixed delay due to the precision loss induced by noise). Second, when normalized by the maximal value, it is a first order approximation of the vector which is especially relevant for over-complete representations where the information contained in the rank vector (which is thanks to Stirling's approximation of order $\log_2(N!)= O(N.\log(N))$, that is more than \unit[2000]{bits} for $256$ neurons) is greater than the information contained in the particular quantization of the image\footnote{We are generally unable to detect quantization errors on an image consisting of more $256$ gray levels, that is for \unit[8]{bits}.}. On a practical note, we may use the fact that the inverse of $f_j$ may be computed from the mean over trials of the function of the absolute functions as a function of the rank. \\
This code therefore focuses on the particular sequence of neurons that were chosen and loses the particular information that may be coded in the pattern of individual inter-spike intervals in the assembly. A model accounting for the exact spiking mechanism would correct this information loss, but this would be at the cost of introducing new parameters (hence new information), while it seems that this information would have a low impact relative to the total information~\citep{Panzeri99}. More generally, one could use different mappings for the transformation of the $z$ value into the a spike volley which can be more adapted to continuous flows, but this scheme corresponds to an extreme case (a transient signal) which is useful to stress on the dynamical part of the coding~\citep{Van-Rullen01a} and is mathematically more tractable. In particular, one may show that the coding error is proportional to the variability of the sorted coefficients~\citep{Perrinet03ieee}, the rest of the information being the information coded in the time intervals between two successive spikes. Thus, the efficiency of information transmission will directly depend on the validity of the hypothesis of independence of the choice of components and therefore on the statistical model build by the LGM.\\%
It should be also noted that no explicit reconstruction is \emph{necessary} (in the mathematical sense of the term) on the receiver side as we do here, since the goal of the receiver could only be to manipulate information on for instance some subset on the spike list (that is on some receptive field covering a subpart of the population). In simple terms, there is no reason to have a reconstruction of the image in the CNS. In particular one may imagine that we may add some arbitrary global point linearity to the $z$ values in order to threshold low values or to quantize values (for instance set all values to $1$ only for the first $10\%$ of the spikes). However, this full reconstruction scheme is a general framework for information transmission, and we may then imagine that if for instance we pool information over a limited receptive field, the information needed (the ranks in the sub-spikelist) will still be available to the receiver directly without having to compute the full set (in fact, since the pdf of $z$ is uniform, the pdf of a subset of components of $z$ is also uniform).%
\section{Sparse Spike Coding}%
\label{sec:SSC}%
However, as we described before~\citep{Perrinet02sparse,Perrinet04tauc,Perrinet06}, if we use over-complete dictionaries of filters, the resulting spiking code gets redundant. In fact, unless the dictionary is orthogonal, when choosing one component over an other, any choice may modify the choice of the other components. If we chose the successive neurons with maximum correlation values, the resulting representation will be proportionally more redundant when the dictionary gets more over-complete. Also, we saw that  optimizing the choice leads then to a combinatorial explosion~\citep{Perrinet08shl}. To solve this NP-complete problem to model realistic representations such as when modeling the primary visual cortex, one may implement a solution designed after the richly laterally connected architecture of cortical layers~\citep{Fischer05a,Fischer06tip,Fischer07cv}. In fact, an important part of cortical areas consists of a lateral network propagating information in parallel between neurons. We will here propose that the NP-problem can be approximately solved by using a cross-correlation based inhibition between neurons.\\%
In fact, as was first proposed in the \emph{Sparse Spike Coding} (SSC) algorithm~\citep{Perrinet02sparse}, one could use a greedy algorithm on the L$_0$-norm cost and that these led to use of Matching Pursuit algorithm~\citep{Mallat93}. More generally, let's first define Weighted Matching Pursuit (WMP) by introducing a non-linearity in the choice step. Like Matching Pursuit, it is based on two repetitive steps. First, given the signal $x$, we are searching for the \textit{single} source $ s^\ast_{j^{\ast}} .  h_{j^\ast}$ that corresponds to the maximum \textit{a posteriori} (MAP) realization for $x$ (see \seeEq{coco}) transformed by a point non-linearity $f_j$. This Matching step is defined by:%
\begin{equation}
 j^\ast = \mbox{ArgMax}_{j} [f_j( \rho_j )]
\label{eq:mp1}
\end{equation}%
where $f_j(.)$ is some gain function that we will describe below and which may be set initially to strictly increasing functions and $\rho_j$ is initialized by Eq.~\ref{eq:coco}. %
In a second step (Pursuit), the information is fed-back to correlated sources through :
\begin{equation}
x \la x - s^\ast_{j^{\ast}} .  h_{j^{\ast}} 
\label{eq:mp2}
\end{equation}
where $s^\ast_{j^{\ast}}$ is the scalar projection $ < x,  h_{j^\ast} > $. Equivalently, from the linearity of the scalar product, we may propagate laterally:
\begin{equation}
 <x, h_j> \la <x, h_j> - < x, h_{j^{\ast}} > < h_{j^{\ast}},  h_j > 
\end{equation}
that is from \seeEq{coco}:
\begin{equation}
\rho_j \la \rho_j - \rho_{j^{\ast}} < h_{j^{\ast}},  h_j > 
\label{eq:mp3}
\end{equation}
For any set of monotonously increasing functions $f_j$, WMP shares many properties with MP, such as the monotonous decrease of the error or the exponential convergence of the coding. The algorithm is then iterated with Eq.~\ref{eq:mp1} until some stopping criteria is reached. The signal may be reconstructed from the spike list as $x = \sum \hat{\rho}_{j(k)} h_{j(k)}$, where $ \hat{\rho}_{j(k)}$ is the value reconstructed using Eq.~\ref{eq:mod}. %
\begin{figure}
\includegraphics[width=.49\textwidth,height=.52\textwidth]{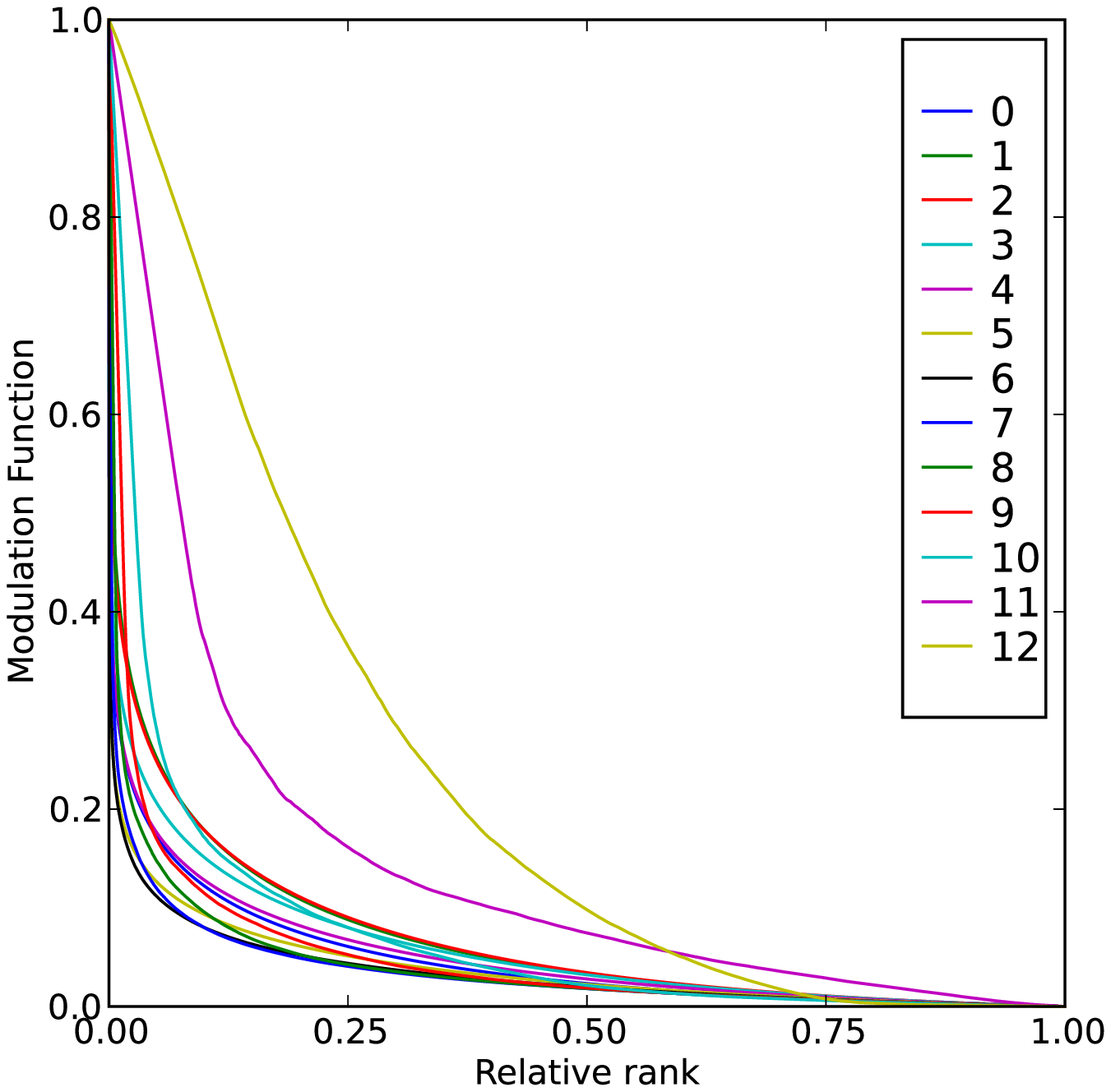}%
\hspace*{.019\textwidth}%
\includegraphics[width=.49\textwidth,height=.475\textwidth]{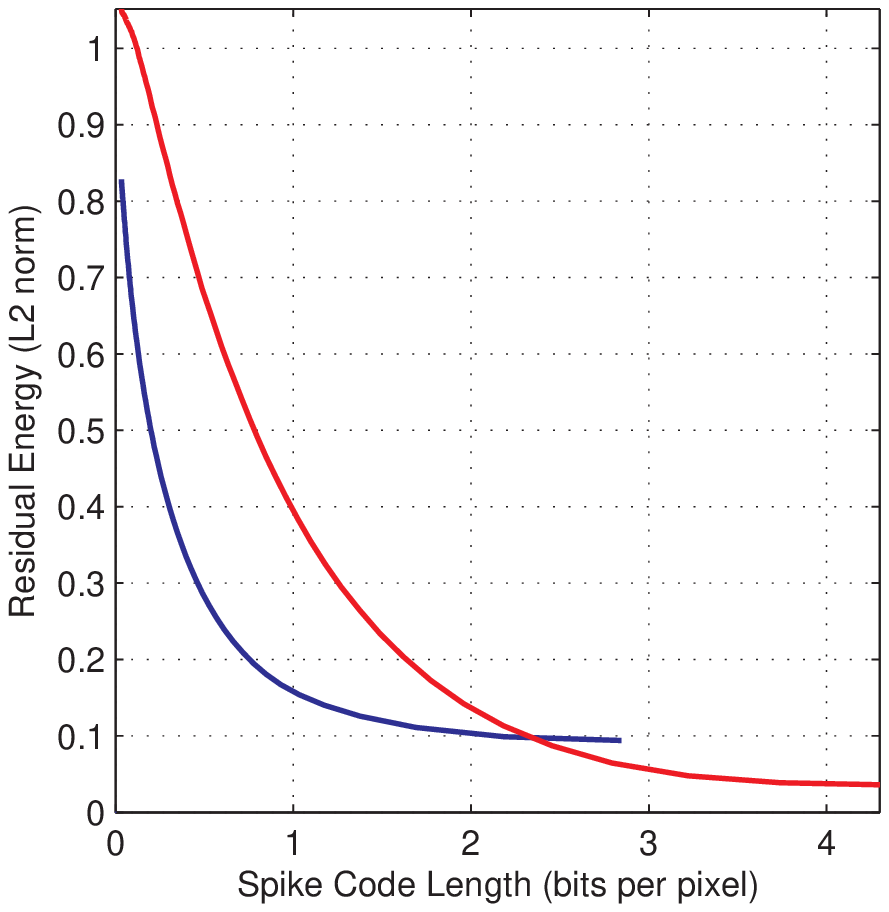}%
\vskip -.47\textwidth 
\hskip .759\textwidth
\includegraphics[width=.22\textwidth]{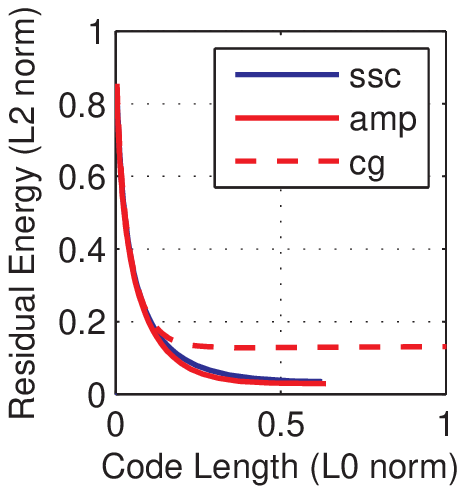}%
\vskip .25\textwidth
   \caption[COMP]
   {\label{fig:COMP} \rm{Efficiency of Competition Optimized Matching Pursuit (COMP). }%
Spike Coding and Sparse Spike Coding (using COMP) produce flows of spikes representing the image. By representing the the distance of the original image with a reconstruction, one may quantify the dynamical efficiency of this solution as a function of the number of spikes. \emph{(Left)} When applying the algorithm on a set of natural images, the coefficients exhibited differences in their probability density functions. We show this by plotting the cumulative density functions of the coefficients for different levels in the pyramid. Using these cumulative pdf, one could transform the pyramids of coefficients in pyramids for which all coefficients where \emph{a priori} equiprobable. This optimizes the ArgMax operator which is at the heart of the Sparse Spike Coding scheme. %
\emph{(Right)} The resulting COMP solution gives a similar result than MP in terms of residual energy as a function of pure $L_0$ sparseness (see inset). In fact, in MP, by taking the maximum absolute, and since the decrease of energy is proportional to the square of the coefficient (see Chapter ~3.1.2 of ~\citep{Perrinet06}) one ensures that the decrease of MSE \emph{per coefficient} is optimal for MP. These are both better for that purpose than conjugate gradient.  However, when defining the efficiency in terms of the residual energy as a function of the description length of the spiking code word, then the proposed COMP model is more efficient than MP because of the quantization errors inherent to the higher variability of coded coefficients. Thus, including homeostasis improved the efficiency of adaptive Sparse Spike Coding by ensuring that the decrease of MSE \emph{per bit of code} is optimal. It should be noted that the homeostasis mechanism is important during ``learning'' but that it is not useful for ``pure'' coding (see Sec.~\ref{sec:SSC}).}%
\label{fig:homeo}%
\end{figure}%
 We then define Competition Optimized Matching Pursuit (COMP) as WMP where the point non-linearities are defined by Eq.~\ref{eq:laughlin} and Sparse Spike Coding (SSC) is then defined as the spike coding/decoding algorithm which uses COMP as the coder. As described in~\citep{Perrinet04tauc}, while the Matching step is efficiently performed by the LIF neurons driven by the NL input, the pursuit step could be implemented in a cortical area by a correlation-based inhibition. This type of inhibition is typical of fast-spiking interneurons though there is no direct evidence of this activity-based synaptic topology. It will correspond to a lateral interaction within the linear (L) neuronal population. In practice, the $f_j$ functions are initialized for all neurons to the identity function (that is to a MP algorithm) and then evaluated using an online stochastic algorithm with a ``learning'' parameter corresponding to a smooth average which effect was controlled. As a matter of fact, this algorithm is circular since the choice of $\sv$ is non-linear and depends on the choice of $f_j$. However, thanks to the exponential convergence of MP, for any set of components, the $f_j$ will converge to the correct non-linear functions as defined by \seeEq{laughlin}. This scheme extends the Matching Pursuit (MP) algorithm by linking it to a statistical model which tunes optimally the matching step (in the sense that all choices are statistically equally probable) thanks to the adaptive point linearity. In fact, as stated before, thanks to the uniform distribution of the choice of a component, one maximizes the entropy of every match and therefore of the computational power of the ArgMax operator. Think \emph{a contrario} to a totally unbalanced network where the match will be always a given neuron: the spikes are totally predictable and the information carried by the spike list then drops to zero. It therefore optimizes the efficiency of MP for the Sparse Spike Coding problem (see Fig.~\ref{fig:SSC}).\\%
Extensions of this type of event-based algorithms are multiple. First, It extends naturally to the temporal domain. In fact, we restricted us ourselves here to static flashed images, but is easily extendable to causal filters (see Ch.~3.4.1 in ~\citep{Perrinet06}). It however raises the unsolved problem of a dynamical compromise between precision and rapidity of the code which is still unanswered. It may also be extended in a adaptive code, showing the emergence of V1-like receptive fields~\citep{Perrinet08shl}. At last, using in these sparse representations of long-range interactions such as those present in the primary visual cortex should prove to be very helpful to resolve generic image processing problems such as denoising.
\subsubsection*{Reproducible science / Acknowledgments}%
All algorithms used in this paper were implemented using Python, Numpy, SciPy (FFT and image libraries) and Matplotlib (for the visualization). Scripts are available upon request. \\%
\subsubsection*{Acknowledgments}%
This work was supported by a grant form the French Research Council  (ANR ``NatStats'') and by EC IP project FP6-015879, "FACETS".%

\end{document}